\documentclass[journal]{IEEEtran}

\usepackage{float}
\usepackage{amssymb}
\usepackage{bm}
\usepackage{amsthm, amsmath,amsfonts}
\usepackage[ruled, vlined]{algorithm2e}
\usepackage{array}
\usepackage{textcomp}
\usepackage{stfloats}
\usepackage{url}
\usepackage{verbatim}
\usepackage{graphicx}
\usepackage{subfigure}
\usepackage{cite}
\usepackage{booktabs}    
\usepackage{mathrsfs} 
\usepackage{setspace}
\usepackage{geometry}
\usepackage{multirow}
\usepackage{amsmath} 
\allowdisplaybreaks[4] 
\usepackage{autobreak} 
\usepackage{xcolor}
\usepackage{hyperref} 
\usepackage{geometry}  
\usepackage{color}
\usepackage{threeparttable}
\makeatletter
\renewcommand{\maketag@@@}[1]{\hbox{\m@th\normalsize\normalfont#1}}%
\makeatother

\begin{document}
\begin{spacing}{1}

\title{
Chirp Delay-Doppler Domain Modulation: A New Paradigm of Integrated Sensing and Communication for Autonomous Vehicles
}
\author{Zhuoran Li, Shufeng Tan, Zhen Gao, \textit{Member}, \textit{IEEE}, Yi Tao, Zhonghuai Wu, Zhongxiang Li, Chun Hu, and Dezhi Zheng
	\thanks{Zhuoran Li, Shufeng Tan, Zhen Gao (corresponding author), Yi Tao, Zhonghuai Wu, Zhongxiang Li, Chun Hu, and Dezhi Zheng are with Beijing Institute of Technology,  China.
	
	}
}


\maketitle
\begin{abstract}	
Autonomous driving is reshaping the way humans travel, with millimeter wave (mmWave) radar playing a crucial role in this transformation to enabe vehicle-to-everything (V2X).
Although chirp is widely used in mmWave radar systems for its strong sensing capabilities, the lack of integrated communication functions in existing systems may limit further advancement of autonomous driving.
In light of this, we first design ``dedicated chirps" tailored for sensing chirp signals in the environment, facilitating the identification of idle time-frequency resources.
Based on these dedicated chirps, we propose a chirp-division multiple access (Chirp-DMA) scheme, enabling multiple pairs of mmWave radar transceivers to perform integrated sensing and communication (ISAC) without interference.
Subsequently, we propose two chirp-based delay-Doppler domain modulation schemes that enable each pair of mmWave radar transceivers to simultaneously sense and communicate within their respective time-frequency resource blocks.
The modulation schemes are based on different multiple-input multiple-output (MIMO) radar schemes: the time division multiplexing (TDM)-based scheme offers higher communication rates, while the Doppler division multiplexing (DDM)-based scheme is suitable for working in a lower signal-to-noise ratio range.
We then validate the effectiveness of the proposed DDM-based scheme through simulations.
Finally, we present some challenges and issues that need to be addressed to advance ISAC in V2X for better autonomous driving.
Simulation codes are provided to reproduce the results in this paper: \href{https://github.com/LiZhuoRan0/2025-IEEE-Network-ChirpDelayDopplerModulationISAC}{https://github.com/LiZhuoRan0}.
\end{abstract}

\section{Overview of Chirp-based Integrated Sensing and Communication}\label{sec_overview}
\IEEEPARstart{F}{or} a long time, chirp waveform has been widely used in radar systems.
Chirp signal is easy to generate, maintains a constant modulus, and facilitate the processing at the receiver.
Consequently, chirp is highly favored for its simplicity in implementation and its low hardware cost\cite{ref_SPM_AutomotiveRadar}.
Recently, chirp-based millimeter-wave (mmWave) radar has exhibited outstanding sensing performance, particularly in autonomous driving applications.
Together with laser imaging, detection, and ranging (LiDAR) and camera, mmWave radar forms one of the three core sensors for autonomous driving.
Chirp-based mmWave radar systems can overcome the limitations of LiDAR and vision systems in poor weather conditions, while also providing superior velocity estimation capabilities\cite{ref_JSTSP_mmWaveRadar}.
Chirp-based autonomous vehicles' sensing results could provide environmental information to support fast vehicle platooning, as well as secure and seamless access for vehicle-to-everything (V2X) communication.

\begin{figure*}[!t]
	\centering
	\color{black}
	\includegraphics[width=7in]{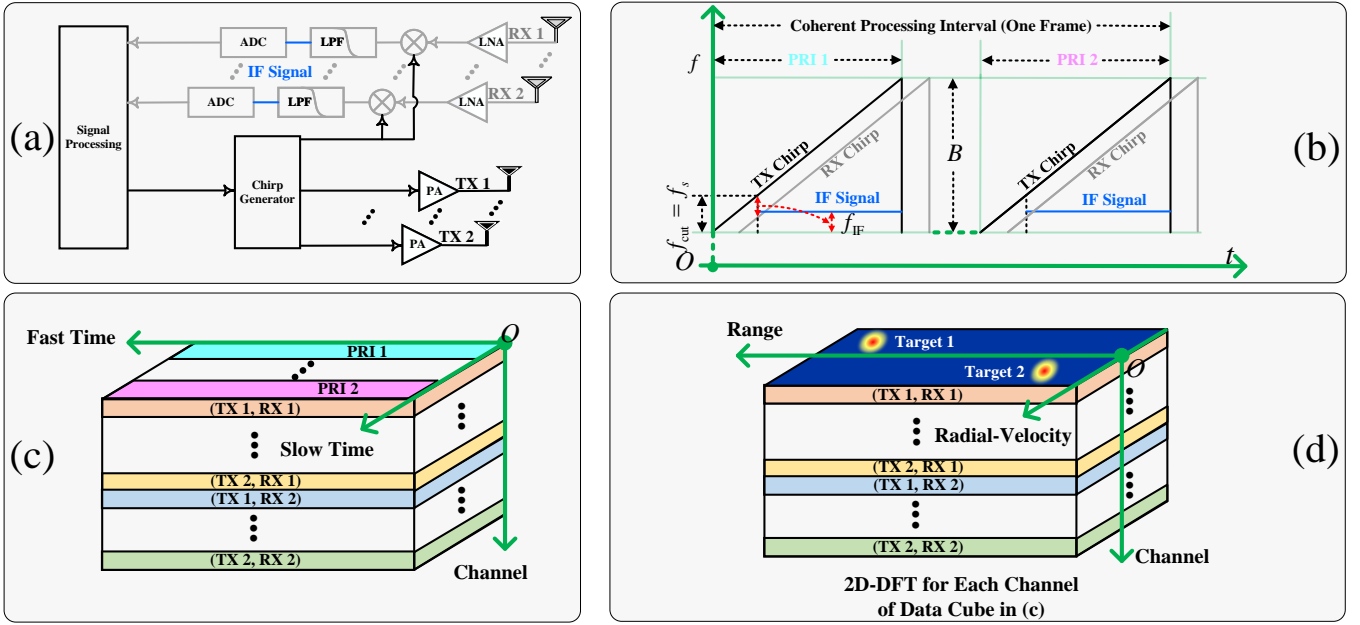}
	\caption{Typical chirp signal processing diagram.
	(a): chirp signal-based transceiver.
	(b): time-frequency diagram of chirp signal with one coherent processing unit (CPI).
	(c): data cube of the receiver.
	(d): range-Doppler map (RDM).
	PRI: pulse repetition interval;
	ADC: analog to digital converter;
	LPF: low-pass filter;
	LNA: low-noise amplifier;
	PA: power amplifier;
	TX: transmit;
	RX: receive.}
	\label{fig_chirpTransceiver}
\end{figure*}

The typical chirp signal processing diagram is shown in Fig.~\ref{fig_chirpTransceiver}.
In Fig.~\ref{fig_chirpTransceiver}\,(a), the chirp from the transmitter's chirp generator serves as the mixing input for the receiver's radio frequency (RF) chain, thereby obtaining the intermediate frequency (IF) signal.
The frequency of the IF signal is $f_{\text{IF}}$, which is much lower than the bandwidth $B$.
$f_{\text{cut}}$ is the cut-off  frequency of the low-pass filter (LPF), while $f_s$ is the sampling frequency of the analog to digital converter (ADC).
For simplicity, we set $f_{\text{cut}}=f_s$.
Generally, $f_{\text{cut}}$ and $f_s$ are in the order of $10$ MHz, while $B$ is in the order of $100$ MHz or even GHz.
Sampling the IF signal for each pulse repetition interval (PRI) yields the data in the fast-time dimension as shown in Fig.~\ref{fig_chirpTransceiver}\,(c).
Stacking multiple PRIs within a coherent processing interval (CPI) forms the data in the slow-time dimension.
Performing two dimensional-discrete Fourier transform (2D-DFT) on the fast-time and slow-time of each ``channel" yields the range-Doppler map (RDM) shown in Fig.~\ref{fig_chirpTransceiver}\,(d), which can be used for estimating distance and velocity.
The ``channels" here are constituted by different transmit and receive antennas.
Combined RDMs from multiple channels can be used for angle estimation.
By employing orthogonal waveforms for each transmit antenna in multiple-input multiple-output (MIMO) radar, the system can obtain $N_{\text{tx}}N_{\text{rx}}$ independent channels for angle estimation\cite{ref_SPM_MIMORadar}.
$N_{\text{tx}}$ and $N_{\text{rx}}$ are the numbers of transmit antennas and receive antennas, respectively.

The success of chirp signals in the radar field has inspired researchers to explore their potential for simultaneous sensing and data transmission.
Since chirp signals were originally developed for radar, related research is often sensing-centric, focusing on scenarios requiring strong sensing capabilities but only moderate communication rates.
This stands in stark contrast to communication-centric waveform solutions based on orthogonal frequency division multiplexing (OFDM)\cite{refTianqiMaoJSAC,ref_WZW,ref_ZYF}.
Although OFDM offers high spectral efficiency, its performance is hindered in the mmWave band by inter-carrier interference (ICI) resulting from high Doppler shifts, which undermines its ability to achieve efficient velocity estimation and maintain reliable communications.
Some studies explored delay-Doppler domain signal processing frameworks like orthogonal time frequency space (OTFS) modulation to enhance Doppler resilience\cite{refTaoJiangIoTJ}, while others aim to combine the advantages of OFDM and chirp, resulting in new waveforms such as orthogonal chirp division multiplexing (OCDM) and affine frequency division multiplexing (AFDM)\cite{ref_2021_ICC_OCDM_AFDM}.
Although these waveforms offer high spectral efficiency and certain sensing capabilities, they come at the cost of more complex hardware and increased computational complexity.
This does not align well with the requirements of autonomous driving, which demands low hardware costs, high sensing capabilities, and low communication rates.

For chirp signals, key parameters include: 1) carrier frequency, 2) bandwidth, 3) PRI, 4) phase, and 5) amplitude.
Thus, some studies proposed to carry data by varying these parameters, such as schemes based on amplitude shift keying (ASK), frequency shift keying (FSK), and phase shift keying (PSK), without altering the chirp spread spectrum (CSS) characteristics of the signal\cite{ref_2013IET_KeWu}.
To maintain the chirp's CSS characteristics and ensure sensing capabilities, the PRI, representing the duration of a single chirp signal, serves as the time unit for data transmission.
Therefore, compared to communication-centric schemes that modulate data at the sample point level, chirp-based schemes inherently face limitations in communication rate.
Depending on the specific requirements, the PRI of chirp can range from a few microseconds to several tens of microseconds, resulting in corresponding communication rates ranging from tens of kilo-bits per second to several hundred kilo-bits per second.

\section{Challenge of Chirp-based Integrated Sensing and Communication}\label{sec_challenge}
Achieving integrated sensing and communication (ISAC) in autonomous driving requires addressing two critical challenges.
The first is interference mitigation: without efficient resource allocation, interference will inevitably occur within the limited time-frequency resources, contradicting the paramount need of safety in autonomous driving.
The second challenge lies in data modulation: since autonomous driving primarily focuses on transmitting control information, an excessive pursuit of high data rates is not cost-effective.
Therefore, we need to carefully balance cost, sensing performance, and communication rate.
\subsection{Interference Mitigation and Resource Allocation}
\begin{figure}[!t]
	\centering
	\color{black}
	\includegraphics[width=3.5in]{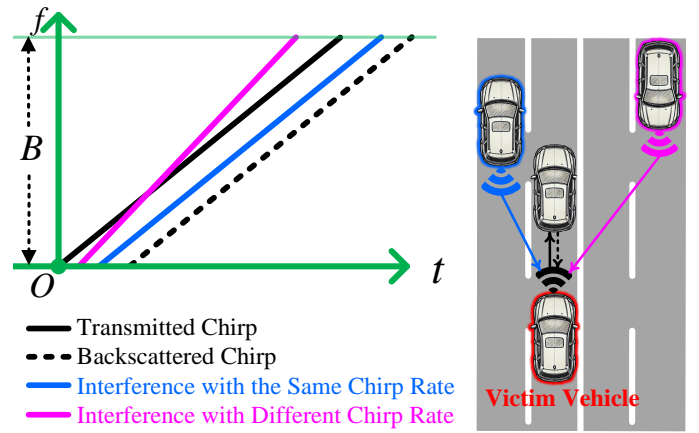}
	\caption{Diagram of different interference sources.}
	\label{fig_interference}
\end{figure}

In autonomous driving scenarios, managing interference is crucial, as interference generally involves one-way transmissions, while useful signals are round-trip.
Consequently, the power of interference could not be underestimated. 
Fig.~\ref{fig_interference} illustrates that, without efficient interference mitigation schemes, the victim vehicle would suffer from severe interference.
The interference originates from chirp signals with the same or different chirp rates whenever there is an overlap in time-frequency resources.
Interference can result in two consequences: the appearance of false targets and an increase in the noise floor.
Therefore, interference mitigation is a crucial prerequisite and guarantee for achieving V2X\cite{ref_2024_Network_V2X}.
To mitigate interference, two strategies can be employed: one is reactive, and the other is proactive\cite{ref_SPM_InterferenceMitigation}.

Existing reactive solutions, involving uncoordinated and random selection of time-frequency resources for transmission, mostly focus on reducing the impact of interference after its occurrence.
These methods include pulse-by-pulse processing to discard corrupted pulses, leveraging frequency diversity, detecting and avoiding occupied frequencies, utilizing narrow main beams, or applying sidelobe null steering\cite{ref_SPM_InterferenceMitigation}.
Utilizing pseudo-random variations in the aforementioned five chirp parameters to counter the coherent accumulation of interference energy is mentioned in \cite{ref_JSTSP_mmWaveRadar}.
Although these solutions can occasionally avoid interference or mitigate minor interference, there is also a significant probability that the interference will be severe and difficult to avoid, posing challenges in meeting the stringent safety requirements of autonomous driving.

For proactive solutions, the probability of interference occurrence can be significantly reduced compared to reactive ones.
Most of these solutions involve initial communication to allocate idle time-frequency resources, followed by sensing with chirp signals.
These solutions require a dedicated communication channel for resource allocation\cite{ref_SPM_InterferenceMitigation}.
However, in small-scale dynamic autonomous driving scenarios, it is feasible to allocate resources autonomously without the need for a central node and dedicated communication frequencies.
If dedicated time-frequency resources are required as a control channel for resource allocation, there is a trade-off between communications and sensing channels in terms of time-frequency resources.
This trade-off indicates that the integration gain has not been fully exploited.
Therefore, cognitive radio can be utilized in coordinated interference mitigation schemes for spectrum coordination.
However, the specific application to mmWave radar in autonomous driving scenarios requires further investigation.
\subsection{Data Modulation based on Chirp Signals}\label{sec_challenge_DataMod}
In the communication realm, chirp is adopted by long range (LoRa) technology for Internet-of-Things (IoT).
LoRa, characterized by its low frequency and narrow bandwidth, is immune to multi-path effects and can correct Doppler frequency offsets through synchronization sequences\cite{ref_Access_ImmuneDoppler}.
The narrow bandwidth of LoRa also facilitates the direct digitization of chirp signals, unlike mmWave radar, where chirp signals are first mixed down to IF before sampling.
Sampling frequency of mmWave radar is significantly smaller than the sweep bandwidth.
Therefore, extending the concept of LoRa to autonomous driving in mmWave band faces three major challenges: multi-path effects, Doppler shifts, and hardware costs.
These factors make this extension difficult.
Although there is an attempt to integrate radar with LoRa, it relies on the nonlinear harmonics generated by RF circuits to produce LoRa signals, significantly restricting its applicability in autonomous driving scenarios\cite{ref_TBC_Loradar}.

Numerous efforts have been made to modulate data without altering the fundamental CSS nature of chirp signals, but many of these efforts focus on single-input single-output (SISO) scenarios, employing ASK, FSK, and PSK to modulate the amplitude, carrier frequency, and phase of chirp signals, respectively\cite{ref_2013IET_KeWu}.
These schemes impose high processing complexity on radar receivers and are challenging to extend to MIMO scenarios.

In MIMO radar systems, orthogonal waveforms are employed for different transmit antennas so that the receiver can distinguish different transmit antennas.
This approach effectively enlarges the aperture of the aperture-constrained array, enhancing angle estimation performance in autonomous driving applications\cite{ref_SPM_MIMORadar}.
For time division multiplexing (TDM)-based scheme, where transmit antennas transmit chirps in turns, the data modulation schemes from SISO scenarios can be directly extended to MIMO scenarios.
However, for non-TDM-based schemes, like Doppler division multiplexing (DDM)-based scheme, where all transmit antennas transmit chirps simultaneously, the received signal is a superposition of all transmit antennas' chirps with different phases.
These phases originate from two sources.
The first is the pre-modulated phases in slow-time.
Specifically, each transmit antenna is pre-modulated with different Doppler shifts, enabling the receiver to distinguish different transmit antennas based on these shifts.
The second source is the different phases caused by the angles of departure.
As a result, the IF signal is no longer a superposition of multi-tone signals, making it impossible to recover communication data PRI-by-PRI.
Moreover, jointly estimating data and sensing parameters from multiple chirps significantly increases the computational complexity.
Therefore, the communication data should be modulated CPI-by-CPI.
There is chirp-wise modulation scheme for MIMO scenarios, where authors of \cite{ref_JSTSP_DingyouMa_IM} utilized selected transmit antennas, carrier frequencies, and phase modulation for information transmission in mmWave MIMO radar systems.
However, it assumed the channel is perfectly known and employed maximum likelihood method for demodulation and estimation, which is challenging to implement due to its high computational complexity.

Given our sensing-centric starting point, with vehicles transmitting only essential control information, we need an ISAC scheme that is simple both in modulation and demodulation.
This scheme should accommodate multiple vehicles simultaneously transmitting signals on the same time-frequency resources without mutual interference, while at the cost of a lower communication rate.

\section{Proposed ISAC Scheme Based on Chirp Signal}\label{sec_scheme}
To address the challenges outlined in Section~\ref{sec_challenge}, we have designed the chirp-division multiple access (Chirp-DMA) scheme for interference mitigation among different transceiver pairs and the chirp-based delay-Doppler domain modulation scheme for ISAC within a pair of transceivers.
The two proposed schemes can significantly enhance the sensing capabilities of autonomous vehicles, promoting a synergistic relationship between their sensing systems and next-generation V2X communication networks.

In the following description of the proposed schemes, we refer to the transceiver that actively transmits ISAC signals and receives the echoes as the active transceiver (AT), while the transceiver that passively receives the ISAC signals is referred to as the passive transceiver (PT).
AT and PT are deployed on two different autonomous vehicles.
Both AT and PT are the same type of transceivers, namely ISAC transceivers, with the only difference being that AT is active and PT is passive.
\subsection{Proposed Chirp-Division Multiple Access}
As shown in Fig.~\ref{fig_chirpTransceiver}, since the received signal of the mmWave radar will go through an LPF after being mixed with the transmitted chirp signal, interference-free ISAC can be achieved as long as there is a certain delay among different chirps.
A similar idea has been proposed in \cite{ref_SPM_InterferenceMitigation}, but it necessitates the use of additional communication resources to allocate time slots.
In fact, by exploiting the chirps' characteristics, we can identify idle time-frequency resources using cognitive radio techniques, without the assistance of a central node.

\begin{figure}[!t]
	\centering
	\color{black}
	\includegraphics[width=3.5in]{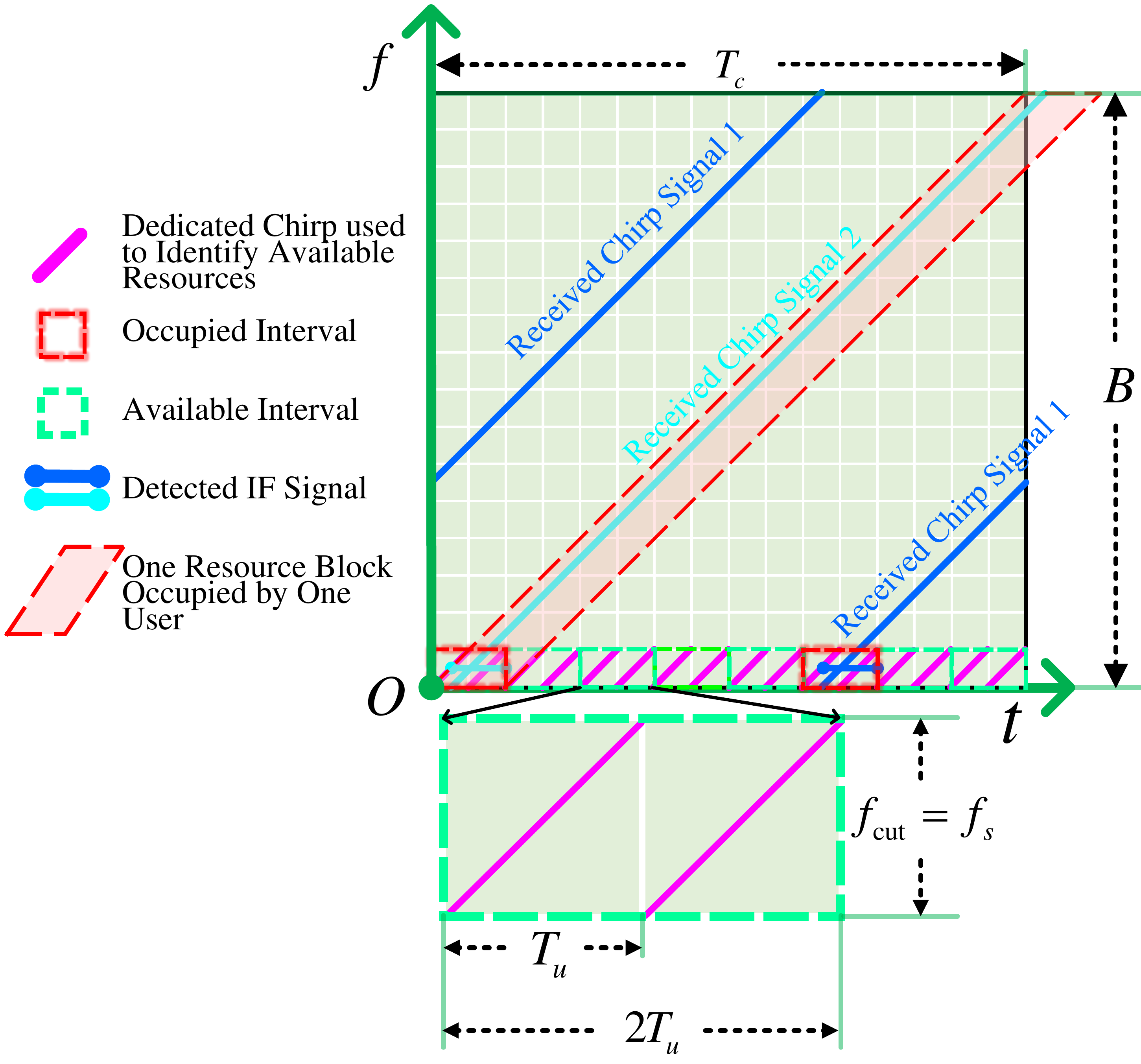}
	\caption{The perception process of available resources using the designed dedicated chirp. ${T_u} = {T_c}{f_{{\text{cut}}}}/B$ is time unit used for ISAC. $2T_u$ is time unit used for Chirp-DMA.}
	\label{fig_ChirpDMA}
\end{figure}

Fig.~\ref{fig_ChirpDMA} shows the perception process of available resources using the designed dedicated chirp.
The dedicated chirp locally generated by the ISAC receiver maintains the same chirp rate as the transmitted chirp, but with a shorter duration. 
The bandwidth swept by the dedicated chirp is $f_{\text{cut}}$, and the duration of the dedicated chirp is ${T_u} = {T_c}{f_{{\text{cut}}}}/B$, where $T_c$ is the duration of one PRI.
Before transmitting ISAC signals, each ISAC transceiver first mixes the received signals with dedicated chirps at the receiver.
If the IF signals appear within certain intervals, it indicates that the tilted bar-like time-frequency resources are occupied in these intervals.
Fig.~\ref{fig_ChirpDMA} illustrates this process, where the ISAC transceiver receives two chirp signals.
The red boxes represent occupied intervals, while the green boxes indicate available intervals.
Each time-frequency resource unit in Fig.~\ref{fig_ChirpDMA}, can not only perform sensing like mmWave radar, but also modulate communication data.
The specific modulation method will be introduced in Section~\ref{sec_DDMod}.
The time difference between the chirp experienced the maximum delay and the subsequent chirp without any delay must be greater than or equal to $T_u$ to avoid interference.
Therefore, an additional time-frequency unit is required to prevent mutual interference.
Consequently, a total of $T_c/(2T_u)$ different transceiver pairs can be accommodated.
Since one resource block occupied by one pair of transceiver is tilted-bar like, we call this multiple access scheme as Chirp-DMA.

\subsection{Proposed Chirp-based Delay-Doppler Domain Modulation}\label{sec_DDMod}
After identifying available resources with the aid of the dedicated chirp, each pair of transceivers can perform interference-free ISAC within its own tilted-bar-like time-frequency resource block.
To minimize the computational and hardware burdens on the receiver, we aim to employ low complexity algorithms for both communication data demodulation and sensing parameters acquisition.
The mmWave radar achieves low overall complexity by adopting low-complexity fast Fourier transform (FFT) for distance and velocity estimations, a mature constant-false-alarm-rate (CFAR) detector for target detection, and an extended Kalman filter (EKF) for target tracking.
The distance and velocity resolution obtained by FFT is acceptable, but the limited antenna aperture results in a low spatial spectrum resolution when only FFT is used.
Therefore, super-resolution algorithms such as multiple signal classification (MUSIC) and estimation of signal parameters via rotational invariance techniques (ESPRIT) are widely utilized\cite{ref_LZR}.
Although super-resolution algorithms increase computational complexity, the relatively small number of antennas in autonomous driving scenarios ensures that the computational burden remains manageable.

\begin{figure*}[!t]
	\centering
	\color{black}
	\includegraphics[width=7in]{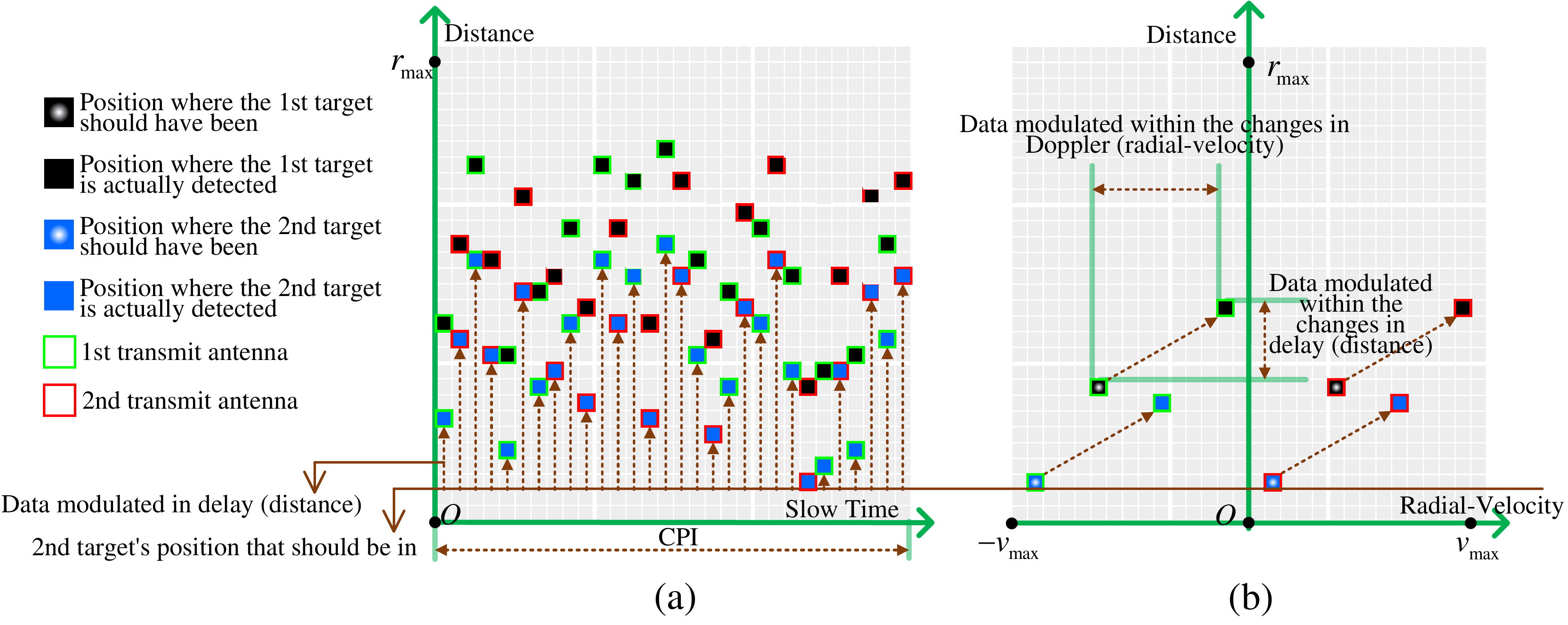}
	\caption{Schematic diagram of data modulation for (a) TDM-based and (b) DDM-based schemes, from the perspective of the communication receiver.
	ISAC transmitter has two transmit antennas and there are two targets in the environment.
	Using the position that the 2nd target should have been in as a reference line is merely for the convenience of plotting and illustrating ``data modulated in delay (distance)'', and the same logic applies to the first target.
	}
	\label{fig_RDM}
\end{figure*}

As shown in Fig.~\ref{fig_chirpTransceiver}\,(c), if data is modulated in the delay-Doppler domain, demodulating communication data does not require additional operations and can be obtained along with sensing parameters.
This is because both of sensing parameters and communication data are obtained after FFT and CFAR detection. 
Communication data is first extracted during the tracking process through the mixture of sensing parameters and communication data.
Tracking is necessary because extracting communication data relies on the prior knowledge of sensing parameters.
Then, the remaining sensing parameters can be used to continue tracking.
The rationale for this approach is that sensing parameters change continuously and slowly, whereas communication data change discretely.
As long as the minimum distance between the designed constellation symbols is larger than the variations in sensing parameters, we can first extract the communication data without being affected by the variations in sensing parameters.

Based on the aforementioned concept, we propose a TDM-based scheme and a DDM-based scheme, both of which modulate data on the variations of delay, Doppler, and complex amplitude.
In TDM-based scheme, delay and complex amplitude can be modulated in each PRI, while Doppler is modulated with CPI as the time unit.
In DDM-based scheme, delay, Doppler, and complex amplitude are all modulated with CPI as the time unit.
The distinction between TDM-based and DDM-based schemes arises from the inherent property of TDM-based schemes that requires antennas to transmit sequentially.
The detailed reasons have been previously introduced in Section~\ref{sec_challenge_DataMod}.
The TDM-based scheme achieves a higher data rate compared to the DDM-based scheme but at the cost of working at higher signal-to-noise ratio (SNR),  because only one antenna is active at a time.
Therefore, different schemes can be adopted according to the specific requirements of the scenario.

Fig.~\ref{fig_RDM} presents the schematic diagram of data modulation for both TDM-based and DDM-based schemes, from the perspective of the communication receiver.
The process shown in Fig.~\ref{fig_RDM} occurs after the initial sensing stage, where the delay and Doppler shift, with no data modulation, are already known.
For the sake of explanation, we define a scenario with two targets in the environment and all ISAC transceivers equipped with two transmit antennas.
Blocks with different colors represent different targets.
The repeated occurrence of blocks of the same color in Fig.~\ref{fig_RDM}\,(b) results from the DDM-based scheme, which enables the receiver to distinguish between transmit antennas by pre-assigning unique Doppler shifts to each\cite{ref_SPM_MIMORadar}.
Fig.~\ref{fig_RDM}\,(b) is RDM, but in the TDM-based scheme shown in Fig.~\ref{fig_RDM}\,(a), the horizontal axis represents the index of PRI within a CPI.
In Fig.~\ref{fig_RDM}\,(a), transmissions from the first antenna are marked by green borders during odd-numbered PRIs, while transmissions from the second antenna are marked by red borders during even-numbered PRIs.
In Fig.~\ref{fig_RDM}\,(b), data modulated in Doppler (radial-velocity) or delay (distance) can be observed from the discrepancies between the target's actual detected position in the RDM and the position it should be in.
In Fig.~\ref{fig_RDM}\,(a), the lengths of different arrows represent the modulated data in the delay domain.
The second target is used as a reference because its smaller delay makes it more convenient for graphical representation.
For the sake of simplicity in the illustration, Doppler modulation is omitted in Fig.~\ref{fig_RDM}\,(a).
In practice, after demodulating the data in the delay-domain, the data modulated in the Doppler-domain can be immediately obtained.

A CPI is defined to have $N_c$ PRIs.
When the minimum modulation spacing in the delay (range) domain is configured to match the delay (range) resolution, each PRI achieves a maximum modulatable capacity of $N_s/2$ distinct points in the delay (range) domain.
Consequently, the maximum detectable round-trip delay (range) of a target is designed as $N_s/2$ range resolution units to maximize the data rate.
The data rate of the TDM-based scheme in the delay (range) domain is ${\left\lfloor {{{\log }_2}\frac{{{T_c}{f_s}}}{2}} \right\rfloor }
/
{{T_c}}$.
Furthermore, each PRI enables complex amplitude modulation on the chirp signal.
The data rate of quadrature amplitude modulation (QAM) or phase modulation (PM) is ${\left\lfloor {{{\log }_2}N_Q} \right\rfloor }
\big/
{{T_c}}$, where $N_Q$ is the order of QAM or PM.
Since data is modulated in Doppler (radial-velocity) once per CPI, the data rate in the Doppler (radial-velocity) domain is ${\left\lfloor {{{\log }_2}N_c} \right\rfloor }
\big/
{(N_c{T_c})}$.
The total data rate of the TDM-based scheme is ${\left\lfloor {{{\log }_2}\frac{{{T_c}{f_s}}}{2}} \right\rfloor }
/
{{T_c}}
+
{\left\lfloor {{{\log }_2}N_Q} \right\rfloor }
/
{{T_c}}
+
{\left\lfloor {{{\log }_2}N_c} \right\rfloor }
\big/
{(N_c{T_c})}
$, which is on the order of hundreds of kilo-bits per second (kbps).
As discussed in Section \ref{sec_challenge}, PRI-by-PRI modulation cannot be utilized in the DDM-based scheme.
The data rates of the DDM-based scheme in the delay (range) domain, in the complex amplitude dimension, and in the Doppler (radial-velocity) domain are ${\left\lfloor {{{\log }_2}\frac{{{T_c}{f_s}}}{2}} \right\rfloor }
/
{(N_c{T_c})}$, ${\left\lfloor {{{\log }_2}N_Q} \right\rfloor }
/
{(N_c{T_c})}$, and ${\left\lfloor {{{\log }_2}N_c} \right\rfloor }
/
{(N_c{T_c})}$ respectively.
The total data rate of the DDM-based scheme is $\left({\left\lfloor {{{\log }_2}\frac{{{T_c}{f_s}}}{2}} \right\rfloor }
+
{\left\lfloor {{{\log }_2}N_Q} \right\rfloor }
+
{\left\lfloor {{{\log }_2}N_c} \right\rfloor }
\right)
/
{(N_c{T_c})}$.
Although the DDM-based scheme achieves only kbps-level rates, this suffices for vehicular control messaging in autonomous driving scenarios.
Additionally, the DDM-based scheme provides SNR advantages over the TDM-based scheme through concurrent multi-antenna transmissions.

\subsection{Performance Evaluations}
\begin{figure*}[!t]
	\color{black}
	\subfigure[]{
		\begin{minipage}[t]{0.33\linewidth}
			\includegraphics[width=2.3in]{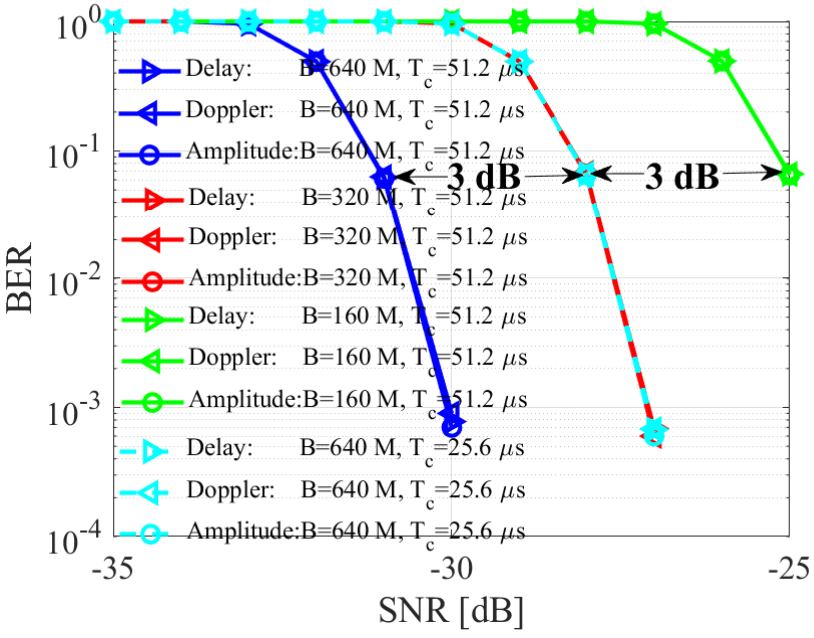}
			\label{fig_BER}
		\end{minipage}%
	}%
	\subfigure[]{
		\begin{minipage}[t]{0.33\linewidth}
			\includegraphics[width=2.3in]{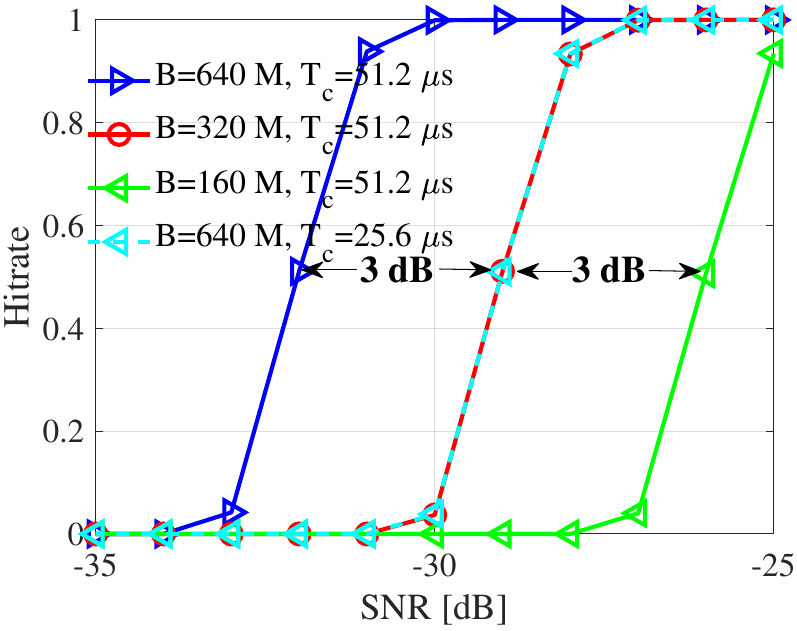}
			\label{fig_Hitrate}
		\end{minipage}%
	}\hspace{-30mm}
	\\
	\subfigure[]{
		\begin{minipage}[t]{0.33\linewidth}
			\includegraphics[width=2.3in]{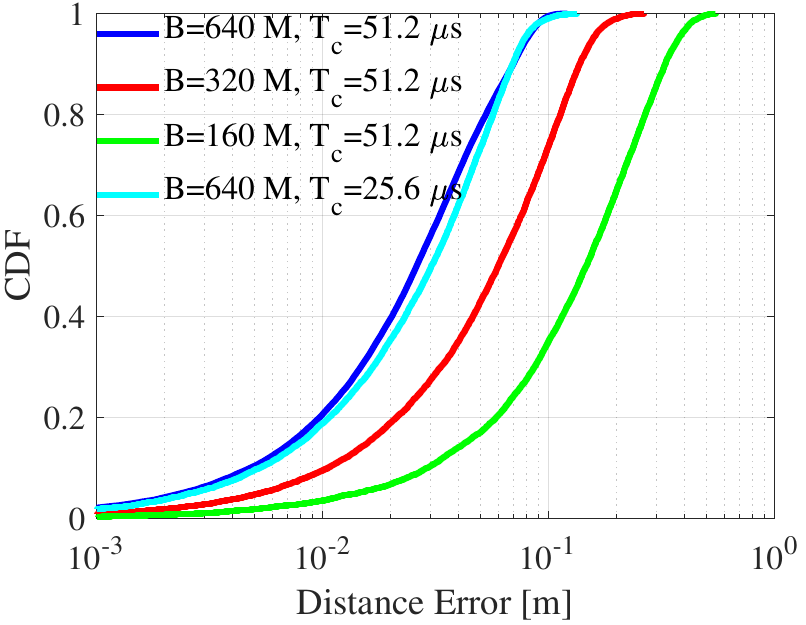}
			\label{fig_CDF_distance}
		\end{minipage}%
	}%
	\subfigure[]{
	\begin{minipage}[t]{0.33\linewidth}
		\includegraphics[width=2.3in]{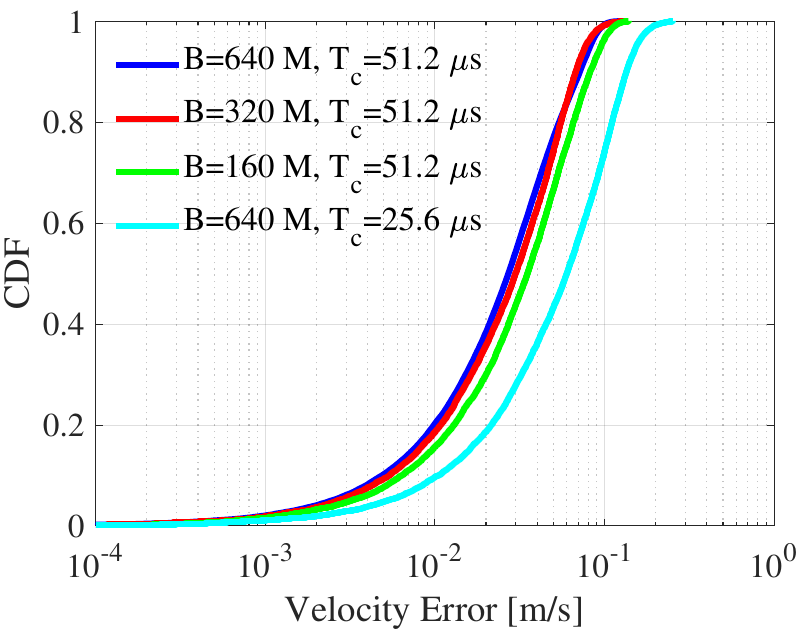}
		\label{fig_CDF_velocity}
	\end{minipage}%
}%
	\subfigure[]{
	\begin{minipage}[t]{0.33\linewidth}
		\includegraphics[width=2.3in]{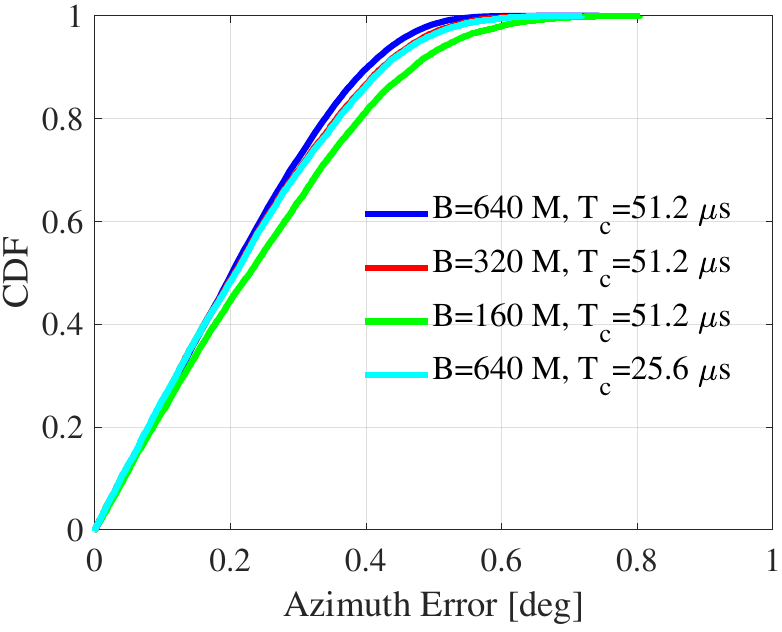}
		\label{fig_CDF_angle}
	\end{minipage}%
}%
	\centering
	\caption{Detection, estimation, and demodulation performance under different parameter settings: bandwidth $B$, PRI $T_c$, and SNR.
		(a): the data demodulation performance of the passive transceiver (PT).
		(b): the detection performance of the active transceiver (AT) using hitrate as performance metric.
		(c): the cumulative distributive function (CDF) of the distance estimation error of the AT.
		(d): the CDF of the velocity estimation error of the AT.
		(e): the CDF of the azimuth angle estimation error of the AT.
		The SNR in (c), (d), and (e) is $-25\,$dB.
	}
	\label{fig_SenComm}
\end{figure*}

The effectiveness of the proposed chirp-based delay-Doppler domain modulation scheme is validated through simulations, with particular focus on its DDM implementation.
The carrier frequency is set to $80$ GHz, with $128$ chirps in one CPI.
The ISAC transceiver has $2\!\times\!2$ (horizontal $\!\!\times\!\!$ elevation) transmit antennas and $2\!\!\times\!\!8$ receive antennas.
The element spacing for the receive antennas is half a wavelength, while the transmit antenna spacing exceeds half a wavelength\cite{ref_SPM_MIMORadar}.
The transmit antenna spacing is related to the number of receive antennas, with the horizontal spacing being four wavelengths and the vertical spacing being one wavelength.
The DDM-based scheme is employed to enable the receiver to distinguish different transmit antennas.
The horizontal field-of-view (FoV) is ($-60^\circ$, $60^\circ$), and the elevation FoV is ($-15^\circ$, $15^\circ$).
The bandwidth can be set to $640\,\text{MHz}$, $320\,\text{MHz}$, or $160\,\text{MHz}$, while the PRI can be set to $51.2\,\mu\text{s}$ or $25.6\,\mu\text{s}$.
We set the ratio of bandwidth $B$ to sampling rate $f_s$ to $32$.
We use QPSK, namely $N_Q=4$.
After the AT and the PT have sensed available time-frequency resources through Chirp-DMA, they perform interference-free ISAC on these time-frequency resources.

We demonstrate the effectiveness of the proposed chirp-based delay-Doppler domain modulation scheme through simulations of the PT's bit-error-rate (BER) performance, the AT's detection performance (characterized by hitrate\cite{ref_JSTSP_DingyouMa_IM}), and the AT's parameter estimation performance (characterized by the cumulative distributive function of the absolute estimiation errors).
A hit happens when the following condition is satisfied: the delay estimation error, velocity estimation error, and angle estimation error are within the corresponding delay resolution, velocity resolution, and $3^\circ$, respectively.
The above condition is flexible and can be adjusted based on specific requirements.
We use cell average-CFAR (CA-CFAR) algorithm \cite{ref_SPM_AutomotiveRadar} after FFT for detection and demodulation, which ensures low computational complexity.
Angle estimation is based on the well-known MUSIC algorithm\cite{ref_LZR}, where the search precision is $1^\circ$.
To improve estimation performance, we aggregate all detected positions caused by spectrum leakage in the RDM and calculate their arithmetic average as the final estimates.
For the sake of simulation convenience and without loss of generality, we set the heights of all vehicles and targets to be the same, resulting in a $0^\circ$ elevation angle.
Therefore, we only show the azimuth estimation error.

The computational complexity of FFT is $\mathcal{O}(N_{\text{rx}}
\times
(N_s N_c \log N_c + N_s N_c \log N_s))$.
The computational complexity of CA-CFAR is $\mathcal{O}(N_{\text{rx}}
\times
N_s
\times
N_c\times
N_{\text{ref}})$, where $N_{\text{ref}}$ is the number of reference units.
We assume one target per range-velocity bin, making the MUSIC algorithm equivalent to the matched filtering with computational complexity $\mathcal{O}(N_{\text{a}}
N_{\text{rx}}
N_{\text{tx}})$, where $N_{\text{a}}$ is the number of angular search grids.
Therefore, the total computational complexity is $\mathcal{O}\left(N_{\text{rx}}
\times
\big(N_s N_c (\log N_c + N_{\text{ref}}) + N_s N_c \log N_s\big)
+
N_{\text{a}}
N_{\text{rx}}
N_{\text{tx}}\right)$.

Fig.~\ref{fig_BER} shows the demodulation performance of the PT.
``Delay" denotes that data is modulated in delay (distance).
``Doppler" denotes that data is modulated in Doppler (radial-velocity).
``Amplitude" denotes that data is modulated in complex amplitude.
When the PRI is fixed and the bandwidth increases, the number of sampling points increases, leading to an improvement in the equivalent SNR.
Therefore, as the bandwidth increases from $160\,$MHz to $320\,$MHz and then to $640\,$MHz, the required SNR for maintaining the same BER decreases by $3\,$dB each time.
When the PRI increases, the equivalent SNR also improves.
Therefore, the BER performance under $B=640\, \text{MHz}, T_c=25.6\,\mu\text{s}$ and $B=320\,\text{MHz}, T_c=51.2\,\mu\text{s}$ configurations is identical.
Additionally, the BER performances for delay, Doppler, and amplitude are identical, indicating a strong correlation between the detection performances of these three parameters.
We can conclude that bandwidth, PRI, and SNR can be interchanged to achieve the same BER performance.
The reason for the worst BER being $1$ instead of $0.5$ is that if the detection fails, no demodulation takes place.

For bandwidth and PRI configurations of 640 MHz and 51.2 $\mu$s, 320 MHz and 51.2 $\mu$s, 160 MHz and 51.2 $\mu$s, and 640 MHz and 25.6 $\mu$s, the corresponding data rates are 2.75 kbps, 2.59 kbps, 2.44 kbps, and 5.19 kbps, respectively.
The low communication rates are due to the use of the DDM-based scheme.
If a TDM-based scheme were used, the data rate could be increased by a factor of $N_c$, typically by two orders of magnitude.
However, signals are transmitted sequentially through four different transmit antennas for the TDM scheme.
As a result, the BER curve for the TDM scheme will shift $6~\text{dB}$ to the right compared to the DDM scheme, under the same bandwidth and PRI configuration.

Fig.~\ref{fig_Hitrate} illustrates the detection performance of the AT with hitrate as the metric.
It can be observed that higher SNR, larger bandwidth, and longer PRI result in better detection performance.
Note that there is a one-to-one correspondence between the AT's detection performance and the PT's demodulation performance.
On the one hand, the calculation of hitrate is based on whether the errors in delay and Doppler are within their respective resolutions.
On the other hand, the minimum distance between the designed constellation symbols is also determined by the delay and Doppler resolution.

Figs.~\ref{fig_CDF_distance}, \ref{fig_CDF_velocity}, \ref{fig_CDF_angle} show the parameter estimation performance of the AT with SNR being $-25\,$dB.
Once detection is successful, the performance of angle estimation remains relatively consistent across different parameter configurations, as the antenna aperture is fixed in these settings.
In contrast, the accuracy of delay and velocity estimation improves significantly with either increased bandwidth or extended observation time. 
This improvement arises from the enhanced delay resolution provided by larger bandwidth and the enhanced velocity resolution enabled by longer observation time.
Due to spectrum leakage, a single target corresponds to multiple values in the RDM.
By applying arithmetic averaging of these values in the RDM to estimate distance and velocity, the resulting estimation errors for both distance and velocity are significantly smaller than the corresponding resolution limits\cite{ref_JSTSP_DingyouMa_IM}.
Similarly, although our angle search precision is $1^\circ$, the actual angle estimation error is less than $1^\circ$.

The proposed scheme can operate at extremely low SNR, thus allowing for lower transmit power.
As a result, it can avoid interfering with other devices sharing the same time–frequency resources.

\section{Open Problems and Future Directions}\label{sec_future}

\subsection{ISAC Signal Processing for Autonomous Vehicles}
The proposed chirp-based delay-Doppler domain modulation is an upgraded version of conventional mmWave radar, employing low-complexity and commonly used radar algorithms: FFT, CFAR, and EKF. 
It is possible to enhance communication and sensing performance without significantly increasing computational complexity by adopting alternative algorithms\cite{ref_ZYF}.
Furthermore, sensing is not solely achieved through mmWave radar.
In fact, autonomous vehicles are equipped with a variety of other sensors.
Integrating sensing results from cameras and LiDARs holds great potential for improving accuracy and reducing overhead\cite{ref_JSTSP_mmWaveRadar}.
\subsection{ISAC Performance Trade-off for Autonomous Vehicles}
Due to our aim for a low complexity implementation of ISAC, both the delay and Doppler resolutions are limited by the DFT.
If constraints on computational complexity are relaxed, the theoretical performances of communication and sensing require further analysis.
This is because a smaller distance between the designed constellation symbols can enhance the communication rate but may also lead to the worse BER.
However, the accuracy of parameter estimation depends on the accuracy of demodulation. 
Therefore, a smaller constellation symbol distance leads to worse parameter estimation performance.
At the same time, larger errors in parameter estimation result in larger tracking errors, which in turn adversely affect the demodulation performance.

\subsection{Multi-Vehicle Networking}
Currently, only one ISAC transceiver on a vehicle is supported for information exchange with one ISAC transceiver on another vehicle.
If multiple ISAC transceivers on a vehicle were engaged in information exchange with multiple ISAC transceivers on other vehicles, selecting the optimal pairings for each transceiver is crucial.
This ensures that ISAC is performed with the minimal transmit power and interference.
Moreover, the sensing capability of a single vehicle or a single transceiver is inherently limited.
With enhanced communication capacity, collaborative sensing among multiple vehicles and multi-sensor systems becomes feasible.
In complex road environments, such multi-vehicle cooperative sensing enables beyond-line-of-sight sensing, significantly enhancing the robustness of sensing systems.

\subsection{Non-Ideal Factors and Safety for  Vehicular ISAC}
We have assumed ideal synchronization, which is impossible in reality.
Therefore, during the sensing stage, it is necessary to compensate the clock drift to better facilitate ISAC.
Moreover, we have assumed that all devices are non-malicious. 
However, given the high safety requirements of autonomous driving, it is crucial to consider how to identify and minimize the impact of malicious interference.
Furthermore, real-world field testing plays a critical role in verifying the robustness and reliability of an ISAC system.
Conducting trials with prototype vehicles on dedicated test tracks or public roads enables researchers and engineers to capture the complexities of actual driving environments.
	
\subsection{Extension to Other Embodied Intelligent Agents}
While this paper focuses on autonomous driving scenarios, it is worth noting that autonomous vehicles can be regarded as a form of embodied intelligent agents, similar to robots or robotic dogs. 
The technology stacks for sensing, motion planning, and control largely overlap across these different types of embodied intelligent systems.
Therefore, the mmWave radar-based ISAC techniques proposed in this paper could be adapted to other robotic platforms.

\section{Conclusion}\label{sec_Conclusion}
Chirp signals currently play a key role in safeguarding autonomous driving.
In this paper, we have enhanced mmWave radar systems by integrating communication functions into chirp signals, advancing both autonomous driving and V2X communication.
First, we have proposed Chirp-DMA to enable interference-free ISAC among multiple pairs of mmWave radar transceivers.
Second, we have proposed chirp-based delay-Doppler domain modulation schemes tailored to two distinct MIMO radar architectures: TDM-based and DDM-based architectures.
These schemes enable each pair of mmWave radar transceivers to achieve simultaneous sensing and communication.
The effectiveness of the chirp-based delay-Doppler domain modulation has been validated through simulations of the sensing and communication capabilities of the DDM-based scheme.
Finally, open problems and potential research directions have been outlined.

\bibliographystyle{IEEEtran}
\bibliography{ref}


\vfill

\end{spacing}
\end{document}